\documentstyle[twoside]{article}

\catcode`\@=11
\long\def\@makefntext#1{ 
\protect\noindent \hbox to 3.2pt {\hskip-.9pt
$^{{\eightrm\@thefnmark}}$\hfil}#1\hfill} 

 \def\@makefnmark{\hbox to 0pt{$^{\@thefnmark}$\hss}}  

\def\ps@myheadings{\let\@mkboth\@gobbletwo
\def\@oddhead{\hbox{} 
\rightmark\hfil\eightrm\thepage}
\def\@oddfoot{}\def\@evenhead{\eightrm\thepage\hfil 
\leftmark\hbox{}}\def\@evenfoot{}
\def\sectionmark##1{}\def\subsectionmark##1{}}

\oddsidemargin=\evensidemargin
\newcounter{sectionc}\newcounter{subsectionc}\newcounter{subsubsectionc}
\renewcommand{\section}[1] {\vspace{12pt}\addtocounter{sectionc}{1}
\setcounter{subsectionc}{0}\setcounter{subsubsectionc}{0}\noindent
	{\tenbf\thesectionc. #1}\par\vspace{5pt}}
\renewcommand{\subsection}[1] {\vspace{12pt}\addtocounter{subsectionc}{1}
	\setcounter{subsubsectionc}{0}\noindent
	{\bf\thesectionc.\thesubsectionc. {\kern1pt \bfit #1}}\par\vspace{5pt}}
\renewcommand{\subsubsection}[1]
{\vspace{12pt}\addtocounter{subsubsectionc}{1}
	\noindent{\tenrm\thesectionc.\thesubsectionc.\thesubsubsectionc.
	{\kern1pt \tenit #1}}\par\vspace{5pt}}
\newcommand{\nonumsection}[1] {\vspace{12pt}\noindent{\tenbf #1}
	\par\vspace{5pt}}

\newcounter{appendixc}
\newcounter{subappendixc}[appendixc]
\newcounter{subsubappendixc}[subappendixc]
\renewcommand{\thesubappendixc}{\Alph{appendixc}.\arabic{subappendixc}}
\renewcommand{\thesubsubappendixc}
	{\Alph{appendixc}.\arabic{subappendixc}.\arabic{subsubappendixc}}

\renewcommand{\appendix}[1] {\vspace{12pt}
        \refstepcounter{appendixc}
        \setcounter{figure}{0}
        \setcounter{table}{0}
        \setcounter{lemma}{0}
        \setcounter{theorem}{0}
        \setcounter{corollary}{0}
        \setcounter{definition}{0}
        \setcounter{equation}{0}
        \renewcommand{\thefigure}{\Alph{appendixc}.\arabic{figure}}
        \renewcommand{\thetable}{\Alph{appendixc}.\arabic{table}}
        \renewcommand{\theappendixc}{\Alph{appendixc}}
        \renewcommand{\thelemma}{\Alph{appendixc}.\arabic{lemma}}
        \renewcommand{\thetheorem}{\Alph{appendixc}.\arabic{theorem}}
        \renewcommand{\thedefinition}{\Alph{appendixc}.\arabic{definition}}
        \renewcommand{\thecorollary}{\Alph{appendixc}.\arabic{corollary}}
        \renewcommand{\theequation}{\Alph{appendixc}.\arabic{equation}}
        \noindent{\tenbf Appendix \theappendixc #1}\par\vspace{5pt}}
\newcommand{\subappendix}[1] {\vspace{12pt}
        \refstepcounter{subappendixc}
        \noindent{\bf Appendix \thesubappendixc. {\kern1pt \bfit #1}}
	\par\vspace{5pt}}
\newcommand{\subsubappendix}[1] {\vspace{12pt}
        \refstepcounter{subsubappendixc}
        \noindent{\rm Appendix \thesubsubappendixc. {\kern1pt \tenit #1}}
	\par\vspace{5pt}}

\newcommand{\textlineskip}{\baselineskip=13pt}
\newcommand{\smalllineskip}{\baselineskip=10pt}

\def\abstracts#1#2#3{{
	\centering{\begin{minipage}{4.5in}\baselineskip=10pt\eightrm
	\centerline{ABSTRACT}
	\parindent=0pt #1\par
	\parindent=15pt #2\par
	\parindent=15pt #3
	\end{minipage} }\par}}

\renewenvironment{thebibliography}[1]			
	{\ninerm
	 \baselineskip=11pt				
	 \begin{list}{\arabic{enumi}.}
	{\usecounter{enumi}\setlength{\parsep}{0pt}
	 \setlength{\leftmargin 17pt}{\rightmargin 0pt}	
	 \setlength{\itemsep}{0pt} \settowidth		
	{\labelwidth}{#1.}\sloppy}}{\end{list}}

\newcounter{itemlistc}
\newcounter{romanlistc}
\newcounter{alphlistc}
\newcounter{arabiclistc}

\newcommand{\fcaption}[1]{
        \refstepcounter{figure}
        \setbox\@tempboxa = \hbox{\eightrm Fig.~\thefigure. #1}
        \ifdim \wd\@tempboxa > 5in
           {\begin{center}
        \parbox{5in}{\eightrm \smalllineskip Fig.~\thefigure. #1 }
            \end{center}}
        \else
             {\begin{center}
             {\eightrm Fig.~\thefigure. #1}
              \end{center}}
        \fi}

\newcommand{\tcaption}[1]{
        \refstepcounter{table}
        \setbox\@tempboxa = \hbox{\eightrm Table~\thetable. #1}
        \ifdim \wd\@tempboxa > 5in
           {\begin{center}
        \parbox{5in}{\eightrm\smalllineskip Table~\thetable. #1 }
            \end{center}}
        \else
             {\begin{center}
             {\eightrm Table~\thetable. #1}
              \end{center}}
        \fi}

\def\@citex[#1]#2{\if@filesw\immediate\write\@auxout	
	{\string\citation{#2}}\fi			
\def\@citea{}\@cite{\@for\@citeb:=#2\do			
	{\@citea\def\@citea{,}\@ifundefined		
	{b@\@citeb}{{\bf ?}\@warning
	{Citation `\@citeb' on page \thepage \space undefined}}
	{\csname b@\@citeb\endcsname}}}{#1}}

\newif\if@cghi
\def\cite{\@cghitrue\@ifnextchar [{\@tempswatrue
	\@citex}{\@tempswafalse\@citex[]}}
\def\citelow{\@cghifalse\@ifnextchar [{\@tempswatrue
	\@citex}{\@tempswafalse\@citex[]}}
\def\@cite#1#2{{$\null^{#1}$\if@tempswa\typeout
	{IJCGA warning: optional citation argument
	ignored: `#2'} \fi}}

\def\pmb#1{\setbox0=\hbox{#1}
	\kern-.025em\copy0\kern-\wd0
	\kern.05em\copy0\kern-\wd0
	\kern-.025em\raise.0433em\box0}

\def\fnt#1#2{\footnotetext{\kern-.3em
	{$^{\mbox{\scriptsize #1}}$}{#2}}}

\def\fpage#1{\begingroup
\voffset=.3in
\thispagestyle{empty}\begin{table}[b]\centerline{\footnotesize #1}
	\end{table}\endgroup}

\def\runninghead#1#2{\pagestyle{myheadings}
\markboth{{\eightit{\quad #1}}\hfill}{\hfill{\eightit{#2\quad}}}}

\font\tenbf=cmbx10
\font\tenit=cmti10
\font\tenit=cmti10
\font\bfit=cmbxti10 at 10pt
 1
 1
 1

\font\ninerm=cmr9

\font\eightrm=cmr8
\font\eightit=cmti8

\def\qed{\hbox{${\vcenter{\vbox{                          
   \hrule height 0.4pt\hbox{\vrule width 0.4pt height 6pt
   \kern5pt\vrule width 0.4pt}\hrule height 0.4pt}}}$}}

\runninghead{Wormholes and Cosmic Strings}
{Wormholes and Cosmic Strings}
\begin{document}
\normalsize\textlineskip
{\thispagestyle{empty}
\setcounter{page}{1}


\vspace*{0.88truein}

\fpage{1}
\centerline{\bf WORMHOLES AND COSMIC STRINGS}
\vspace*{0.035truein}
\centerline{}
\vspace{0.37truein}
\centerline{\footnotesize PEDRO F. GONZ\'ALEZ-D\'IAZ}
\vspace*{0.015truein}
\centerline{\footnotesize\it Instituto de Matem\'aticas y F\'{\i}sica
Fundamental, Consejo Superior de Investigaciones Cient\'{\i}ficas}
\baselineskip=10pt
\centerline{\footnotesize\it Serrano 121, 28006 Madrid, Spain}
\vglue 10pt

\vspace{0.225truein}


\vspace*{0.21truein}
\abstracts{\noindent A gravitational scenario is proposed where the euclidean
action
is invariant under the isotropic and homogeneous version of the euclidean
$U(1)$ group of local transformations of the scale
factor and scalar matter field, interpreting the trace of the second
fundamental
form as the gauge field.
The model allows spontaneous breakdown of some involved symmetries,
including possibly diffeomorphism invariance, and leads to the
formation of topological deffects. In particular, we consider here
the case of the wormhole-induced formation of cosmic strings.}{}{}
\vspace*{13pt}\textlineskip

\section{Introduction}

This report proposes a possible framework where canonical theory of gravity
coupled to matter can be interpreted as a gauge theory that may undergo
spontaneous symmetry breaking. More precisely, we show that in a
Robertson-Walker minisuperspace with scale factor $a$ and conformal time
$\eta=\int\frac{d\tau}{a}$, the euclidean action for Hilbert-Eisntein
gravity conformally coupled to a scalar field $\Phi$ with generally
nonzero mass $m$ satisfies an "axionic" symmetry that allows the action
to become also invariant under the isotropic and homogeneous version of
the euclidean $U(1)$ group of local transformations, provided we Wick
rotate anti-clockwise and interpret the trace of the second fundamental
form as the gauge field. The final lagrangian can be seen to have the
form of a typical Higgs model, and spontaneous symmetry breaking provides
the resulting baby universes with the maximal Higgs masses. We also
show that in the resulting gauge theory the nucleation of baby
universes can be regarded as a topological process leading to
the formation of well known topological deffects, such as cosmic
strings. One could be tempted to suggest the generalization that
there is a strong link between the topological changes leading
to nucleation of baby universes in quantum gravity, and the
emergence of all kinds of topological deffects that are allowed
in some gauge theories with spontaneous symmetry breaking.

\section{A Wormhole-Higgs Model}

Using ordinary (matter-field) euclidean rotation, $t\rightarrow -i\tau$,
the euclidean Hilbert-Einstein action for a conformally coupled
scalar field in Robertson-Walker metric becomes
\begin{equation}
I=\frac{1}{2}\int d\eta N(-\frac{a'^{2}}{N^{2}}+\frac{\chi
'^{2}}{N^{2}}-a^{2}+\chi^{2}+m^{2}\chi^{2}a^{2}),
\end{equation}
where $N$ is the lapse function, $\chi=(\frac{4\pi G}{3})^{\frac{1}{2}}a\Phi$,
and a prime denotes differentiation with respect to $\eta$. The equations
of motion for $\chi$ and $a$ derived from (1), $\chi ''=\chi+m^{2}a^{2}\chi$
and $a''=a-m^{2}\chi^{2}a$, should correspond to classical solutions
that allow the effective gravitational constant $G_{eff}=G(1-\frac{4\pi
G}{3}\Phi^{2})^{-1}$
to change sign along the allowed phase-space region, and this would ultimately
lead to negative energies and instabilities for perturbations about the
classical solutions$^{1}$. This problem points to inadequacy of the naive
insistence of a single general Wick rotation both for gravity and matter
fields$^{2}$, and suggests thereby rotation in the opposite direction$^{3}$,
$t\rightarrow +i\tau$, as the "correct" euclidean continuation when the
coupled field is constant or pure gravity is dealt with.

Although (1) is invariant under the global transformation (rotation)
\[\Phi\rightarrow\Phi e^{i\alpha_{0}}\]
\[a\rightarrow ae^{-i\alpha_{0}},\]
with $\alpha_{0}$ a constant, it is not under the Robertson-Walker version
of the corresponding local transformation where
$\alpha_{0}\rightarrow\alpha\equiv\alpha(\eta)$,
unless asymptotically. Thus, applying transformation $a\rightarrow
ae^{-i\alpha(\eta)}$,
$\chi\rightarrow\chi$ to action (1), this is amounted with a new term
$\bigtriangleup I^{(\alpha)}=\frac{1}{2}\int d\eta Na^{2}\alpha '^{2}$,
and from the equation of motion for $\alpha$ we obtain $\alpha
'=\frac{B_{0}}{a^{2}}$,
where $B_{0}$ is a constant. It follows $\bigtriangleup
I^{(\alpha)}=\frac{1}{2}\int d\eta NB_{0}^{2}/a^{2}$
which vanishes as $a\rightarrow\infty$. Note nevertheless that the set of the
above two equations of motion for $\chi$ and $a$, as well as the hamiltonian
constraint, $\frac{\delta I}{\delta N}$, and the action itself, remain all
unchanged under the symmetry $\chi=ia$. It turns out that such a symmetry
implies constant imaginary field $\Phi=i(\frac{3}{4\pi G})^{\frac{1}{2}}$,
i.e. a constant axionic field, and this validates rotation of time in the
opposite direction$^{3}$, and
should require introducing an
additional boundary term$^{4}$, $I_{B}=Const.N\int_{0}^{\infty}d\eta$ in
the euclidean action for transitions with constant $\Phi$. Consistency of
the equations of motion and hamiltonian constraint with $\chi=ia$ leads
then to $Const.=R_{0}^{2}$, with $R_{0}^{2}$ the integration constant for
the first-integrated equations of motion which, after symmetry $\chi=ia$,
can be written for the scale factor (in the gauge $N=1$)
\begin{equation}
a'^{2}=a^{2}+\frac{1}{2}m^{2}a^{4}-R_{0}^{2},
\end{equation}
whose classical solution in terms of the Robertson-Walker time $\tau$ is
\begin{equation}
a(\tau)=m^{-1}[(1+2m^{2}R_{0}^{2})^{\frac{1}{2}}\cosh(2^{\frac{1}{2}}m\tau)-1]^{\frac{1}{2}}; \chi=ia,
\end{equation}
which represents a nonsingular wormhole spacetime. The lorentzian continuation
of (3) describes a baby universe with maximum size of the order (for very
small $m$) $R_{0}$, with a singularity in the past at lorentzian time
$t\simeq R_{0}$. Of course, for $m=0$ we recover the usual Tolman-Hawking
wormhole$^{1}$ $a=R_{0}\cosh\eta$, though still satisfying $\chi=ia$.

It is worth pointing out that if symmetry $\chi=ia$ holds and euclidean
rotation is performed anti-clockwise, then remarkably the resulting
euclidean action becomes positive-definite and hence free from the fatal
conformal divergences which have many times been considered as one of
the biggest problems in euclidean gravity.

Re-expressing (1) in terms of $\chi$ alone by using $\chi=ia$, after Wick
rotating in the opposite direction, one can obtain the lagrangian density
\[{\bf
L}=\frac{1}{2}(\frac{\dot{a}}{a})^{2}\Phi^{2}+(\frac{\dot{a}}{a})\Phi\dot{\Phi}+\frac{1}{2}\dot{\Phi}^{2}\]
\begin{equation}
-\frac{1}{2}\frac{\Phi^{2}}{a^{2}}-\frac{1}{4}(\frac{m}{m_{p}})^{2}\Phi^{4}+\frac{m_{p}^{2}R_{0}^{2}}{2a^{4}},
\end{equation}
where the overhead dot means differentiation with respect to $\tau$, $m_{p}$
denotes Planck mass and, if all symmetries are preserved, $\dot{\Phi}=0$.

It is only when the lagrangian density is expressed in the $\chi$-saturated
form (4) that it becomes invariant under the above local transformations.
Denoting $A=\frac{TrK}{3Q}$, where $TrK$ is the trace of the second
fundamental form, $TrK=\frac{3\dot{a}}{a}$, and $Q$ is the charge of the
scalar field, such  transformations can inmediately be written in the more
familiar form
\begin{equation}
\Phi\rightarrow\Phi e^{i\alpha(\tau)}
\end{equation}
\begin{equation}
A\rightarrow A-i\dot{\alpha}(\tau),
\end{equation}
which exactly are the transformations of the isotropic and homogeneous
euclidean ($t\rightarrow +i\tau$) version of the Abelian group $U(1)$ if
$A$ is interpreted as the gauge field. Thus, if $\Phi$ is shifted by some
real scalar $\varphi$, $\Phi\rightarrow\phi=\Phi+\varphi$, then upon
substitution of $\Phi$ for $\phi$, (4) becomes a typical Higgs model
for an arbitrary charge-$Q$ field $\phi$, a massless gauge field $A\propto K$,
and variable "tachyonic" mass $\frac{1}{a}$. Without loss of generality
one may regard that after spontaneous symmetry breaking it is only the
originally zero real component of the scalar field that acquires some
constant classical part. Actually, a full euclidean version of the usual
isotropic and homogeneous Higgs mechanism for spontaneous symmetry breaking
in the $U(1)$ group can be readily obtained. Therefore, after symmetry breaking
in the unitary gauge, instead of the massless gauge scalar field $A$ and the
axion field $\Phi$ one deals with the massive scalar field with mass
$m_{A}=\frac{m_{p}}{ma}$,
and with the real scalar field $\varphi$ with the mass
$m_{\varphi}=(\frac{2}{a^{2}})^{\frac{1}{2}}$.
On the asymptotic region, $\eta\rightarrow\infty$,
$A=\frac{1}{3}(\frac{m^{2}}{2Q^{2}})^{\frac{1}{2}}$
and $m_{A}=m_{\varphi}=0$, so the symmetry is restored on that region. The
most stable broken phase condenses just at the wormhole neck where $A$ vanishes
and the field masses become maximum, $m_{A}=\frac{Qm_{p}}{mR_{0}}$,
$m_{\varphi}=(\frac{2}{R_{0}^{2}})^{\frac{1}{2}}$, so one should expect
the branched baby universes to carry such maximum masses generated by the
Higgs mechanism.

The masses generated in the above Higgs mechanism would provide the
four-momentum constraints with nonzero r.h.s. terms, and it is in this
sense that such a mechanism can be thought to break diffeomorphism
invariance$^{5}$. In fact, Goldstone bosons appearing in the process
could be regarded as gravitons arising from spontaneous breakdown of
diffeomorphism invariance$^{6-8}$, and the trace of the second fundamental
form -i.e. the gauge field- would play the role of the cosmological
time$^{9}$ needed to endow the Wheeler DeWitt equation with a r.h.s.
energy operator.
The gravitationally-induced Higgs mechanism discussed in this report suggests
the possibility that baby universes carrying all the gauge mass of the
broken phase be nucleated. For finite gauge coupling, masses $m_{A}$ and
$m_{\varphi}$ will take on nonzero arbitrarily large values inside the
baby universes, but become exactly zero on the asymptotic region where
the symmetry is restored.

\section{Gravitational Cosmic Strings}

The gauge theory we have considered so far allows for spontaneous
symmetry breaking, and the quantum-gravitational topological
changes that lead to the nucleation of baby universes can
be viewed as phase transitions where topological deffects
are produced. Such deffects would be expected to manifest
as e.g. euclidean cosmic strings, with static string
metric$^{10}$
\begin{equation}
ds^{2}=e^{2\nu}d\tau^{2}+e^{2\psi}d\phi^{2}+e^{2\lambda}(d\rho^{2}+dz^{2}),
\end{equation}
where we have chosen the string to lie along the $z$ axis, and
$\nu$, $\psi$ and $\lambda$ will now depend not just on the
radial internal coordinate $\rho$, but also on the scale factor
$a$, and $\phi=0$ and $\phi=2\pi$ are identified. Taking for the
interior of the string a uniform energy density $\epsilon(a)>0$,
out to some cylindrical radius $\rho_{0}$, guarantees
avoidance of any singularity if the transverse dimensions of
the strings are comparable to the Planck length, and we assume
$R_{0}>>m_{p}^{-1}$, $mR_{0}<<1$. One can then use the thin-string
approximation$^{10}$. Now, since we must choose
\[\epsilon=\frac{m_{p}}{2\pi^{2}ma^{4}}, \rho<\rho_{0},\]
which is independent of $\rho$, and all components of the
stress-tensor equal to zero, unless $T_{\tau}^{\tau}=T_{z}^{z}=-\epsilon$,
from the Einstein equations for metric (7), we obtain an interior
metric
\begin{equation}
ds^{2}=d\tau^{2}+d\rho^{2}+dz^{2}+m^{*2}a^{4}\sin^{2}(\frac{\rho}{m^{*}a^{2}})d\phi^{2},
\end{equation}
where $m^{*2}=\frac{\pi m}{4Gm_{p}}$, and $a$ is given by (3).

Metric (8) differs from the Hiscock internal metric$^{10}$ by
its dependence on the scale factor $a$; they only become the same
at the wormhole neck, where $m^{*}a^{2}=(8\pi G\epsilon_{0})^{-\frac{1}{2}}$,
$\epsilon_{0}=\frac{m_{p}}{2\pi^{2}mR_{0}^{4}}$. On the
asymptotic region, $a\rightarrow\infty$, metric (8) reduces to
an euclidean cylinder of radius $\rho$, so the interior of the
string disappears. Following the same procedure as in Ref. 10,
we also obtain an euclidean static, cylindrically symmetric
exterior metric
\begin{equation}
ds^{2}=d\tau^{2}+dz^{2}+dr^{2}+(1-4\mu(a))^{2}r^{2}d\phi^{2},
\end{equation}
where the mass per unit length will now depend on
the wormhole scale factor
\[\mu(a)=\frac{1}{4}[1-\cos(\frac{\rho_{0}}{m^{*}a^{2}})].\]
At the wormhole neck, (9) reduces to the euclidean version
of the Vilenkin conical metric$^{11}$. Asymptotically,
$\mu\rightarrow 0$ and the string closes upon itseft; i.e.
on the asymptotic region there is neither exterior nor
interior of strings: topological deffects simply disappear.
Rotating to the lorentzian region, $\tau\rightarrow -it$,
metrics (8) and (9), it can be seen that they become
those of a usual cosmic string$^{10,11}$ only when baby
universes reach their maximum radius $R_{0}$.

\nonumsection{Acknowledgements}
\noindent
This work was supported by a CAICYT Research Project N§ 91-0052.

\nonumsection{References}

\end{document}